\documentclass[superscriptaddress,twocolumn,prl,preprintnumbers,amsmath,amssymb]{revtex4}
\usepackage{graphicx,afterpage}
\begin{document}

\title{Observation of electronic bound states in charge-ordered YBa$_2$Cu$_3$O$_y$}

\author{R. Zhou}
\affiliation{Laboratoire National des Champs Magn\'etiques Intenses, CNRS - Universit\'e Grenoble Alpes - UPS - INSA - EMFL, 38042 Grenoble, France}
\author{M. Hirata}
\altaffiliation[Present address: ]{Institute for Materials Research (IMR), Tohoku University, Katahira 2-1-1, Aoba-ku, Sendai 980-8577, Japan}
\affiliation{Laboratoire National des Champs Magn\'etiques Intenses, CNRS - Universit\'e Grenoble Alpes - UPS - INSA - EMFL, 38042 Grenoble, France}
\author{T. Wu}
\altaffiliation[Present address: ]{Hefei National Laboratory for Physical Sciences at the Microscale, University of Science and Technology of China (USTC), Anhui, Hefei 230026, P. R. China}
\affiliation{Laboratoire National des Champs Magn\'etiques Intenses, CNRS - Universit\'e Grenoble Alpes - UPS - INSA - EMFL, 38042 Grenoble, France}
\author{I. Vinograd}
\affiliation{Laboratoire National des Champs Magn\'etiques Intenses, CNRS - Universit\'e Grenoble Alpes - UPS - INSA - EMFL, 38042 Grenoble, France}
\author{H. Mayaffre}
\affiliation{Laboratoire National des Champs Magn\'etiques Intenses, CNRS - Universit\'e Grenoble Alpes - UPS - INSA - EMFL, 38042 Grenoble, France}
\author{S. Kr\"amer}
\affiliation{Laboratoire National des Champs Magn\'etiques Intenses, CNRS - Universit\'e Grenoble Alpes - UPS - INSA - EMFL, 38042 Grenoble, France}
\author{M. Horvati\'c}
\affiliation{Laboratoire National des Champs Magn\'etiques Intenses, CNRS - Universit\'e Grenoble Alpes - UPS - INSA - EMFL, 38042 Grenoble, France}
\author{C.Berthier}
\affiliation{Laboratoire National des Champs Magn\'etiques Intenses, CNRS - Universit\'e Grenoble Alpes - UPS - INSA - EMFL, 38042 Grenoble, France}
\author{A.P. Reyes}
\affiliation{National High Magnetic Field Laboratory, Florida State University, Tallahassee, Florida 32310, USA}
\author{P.L. Kuhns}
\affiliation{National High Magnetic Field Laboratory, Florida State University, Tallahassee, Florida 32310, USA}
\author{R. Liang}
\affiliation{Department of Physics and Astronomy, University of British Columbia, Vancouver, BC, Canada, V6T~1Z1}
\affiliation{Canadian Institute for Advanced Research, Toronto, Canada}
\author{W.N.~Hardy}
\affiliation{Department of Physics and Astronomy, University of British Columbia, Vancouver, BC, Canada, V6T~1Z1}
\affiliation{Canadian Institute for Advanced Research, Toronto, Canada}
\author{D.A.~Bonn}
\affiliation{Department of Physics and Astronomy, University of British Columbia, Vancouver, BC, Canada, V6T~1Z1}
\affiliation{Canadian Institute for Advanced Research, Toronto, Canada}
\author{M.-H. Julien}
\email{marc-henri.julien@lncmi.cnrs.fr}
\affiliation{Laboratoire National des Champs Magn\'etiques Intenses, CNRS - Universit\'e Grenoble Alpes - EMFL, 38042 Grenoble, France}
\date{\today}

\pacs{74.25.nj, 74.72.Gh, 71.45.Lr, 74.62.En, 76.60.Cq}

\begin{abstract}

Observing how electronic states in solids react to a local symmetry breaking provides insight into their microscopic nature. A striking example is the formation of bound states when quasiparticles are scattered off defects. This is known to occur, under specific circumstances, in some metals and superconductors but not, in general, in the charge-density-wave (CDW) state. Here, we report the unforeseen observation of bound states when a magnetic field quenches superconductivity and induces long-range CDW order in YBa$_2$Cu$_3$O$_y$. Bound states indeed produce an inhomogeneous pattern of the local density of states $N(E_F)$ that leads to a skewed distribution of Knight shifts which is detected here through an asymmetric profile of $^{17}$O NMR lines. We argue that the effect arises most likely from scattering off defects in the CDW state, which provides a novel case of disorder-induced bound states in a condensed-matter system and an insightful window into charge ordering in the cuprates.

\end{abstract}
\maketitle

Recent experiments in high-$T_c$ copper-oxides at moderate doping levels have led to a consensus regarding the presence of static, but essentially short-range, spatial modulations of the charge density that appear far above the critical temperature $T_c$ and compete with superconductivity below $T_c$~\cite{Wu15,Comin16,Ghiringhelli12,Achkar12,Chang12,Croft14,Thampy14,Tabis14,daSilvaNeto15,daSilvaNeto14,Hashimoto14,Comin14}. As a result, the debate has mainly focused on this "incipient" charge-density-wave (CDW) in zero magnetic field.  Somewhat less attention has been paid to another CDW order that also coexists and competes with superconductivity but is present at lower temperatures, only in high magnetic fields, and has been detected only in YBa$_2$Cu$_3$O$_y$ (YBCO) so far, by nuclear magnetic resonance (NMR)~\cite{Wu11,Wu13} and more recently by X-ray diffraction~\cite{Gerber15,Chang16}. This CDW develops long-range three-dimensional (3D) correlations above a critical field (also detected in sound velocity~\cite{LeBoeuf13} and thermal conductivity measurements~\cite{Grissonnanche15}). In principle, the long-range coherence of the high-field CDW makes it a unique playground for investigating the nature of CDW order in cuprates and its interplay with superconductivity~\cite{Julien15}. However, since its observation requires high fields, experimental characterization of this phase has remained limited. This evidently prompts for further high field experiments in YBCO.

Here, we report that a spatial distribution of $^{17}$O Knight shift values develops alongside with the high-field CDW in YBCO. After discarding broadening by the vortex lattice or magnetic order, we explain the effect by an enhancement of $N(E_F)$, the {\it local} density-of-states (LDOS) at the Fermi energy $E_F$, due to the formation of quasiparticle bound states as observed around defects in the superconducting state of cuprates~\cite{Ouazi06,Harter07,Hudson99,Yazdani99,Pan00,Chatterjee08,Balatsky06,Alloul09}. This unforeseen observation raises fundamental questions on the electronic band structure and the nature of defects in the CDW state.

\begin{figure*}[t!]
\centerline{\includegraphics[width=6in]{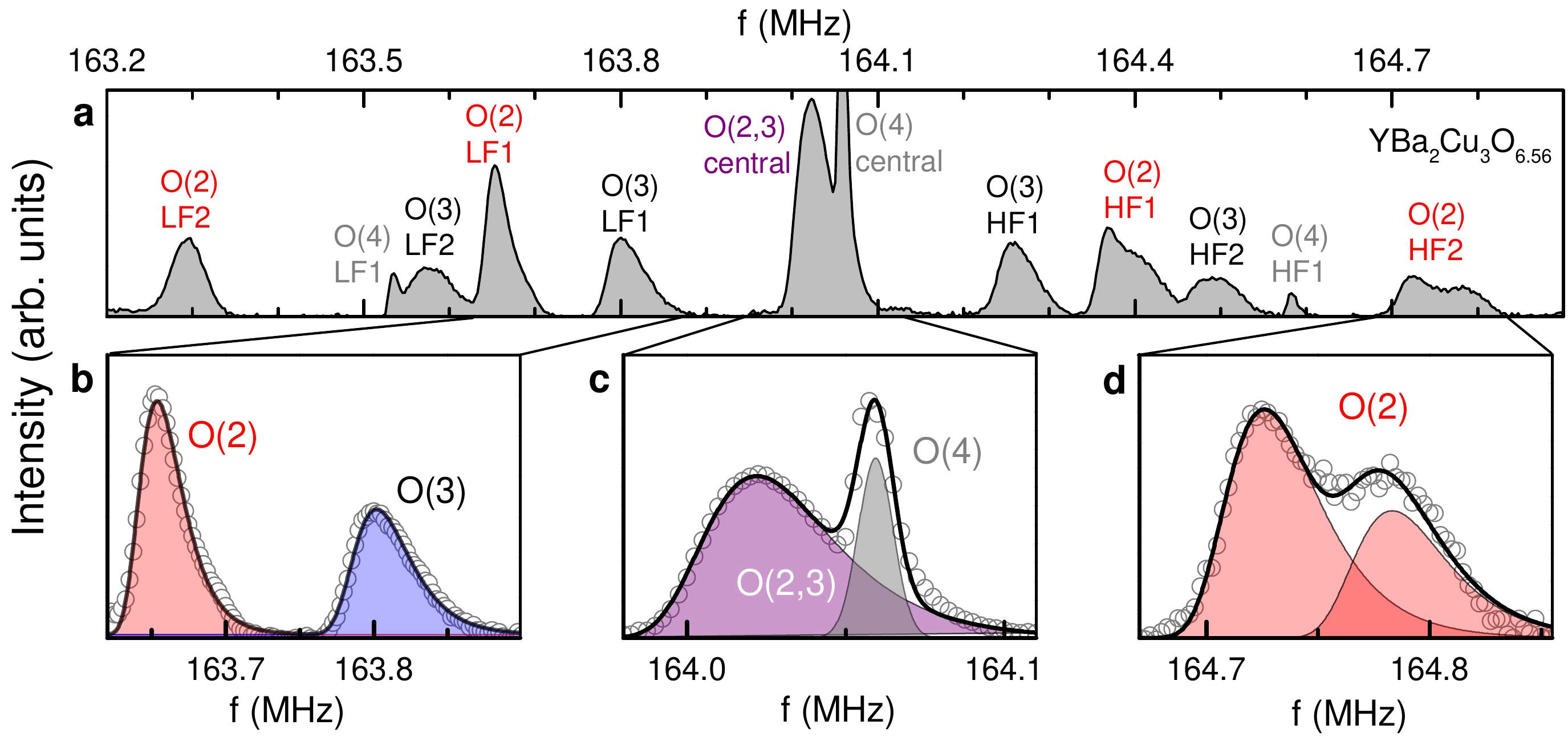}} 
 \caption{(Color online). (a) $^{17}$O NMR spectrum of YBa$_2$Cu$_3$O$_{6.56}$ at $T=3$~K and $H\simeq28.5$~T (27.4~T for the $c$-axis projection as the field is tilted off the $c$-axis~\cite{suppl}). HF1(2) are the first (second) high-frequency satellites and LF1(2) are the first (second) low-frequency satellites. For the apical O(4) site, HF2 and LF2 lines are out-of-scale. (b-d) Zoom on particular lines with fits using an extreme value distribution function (see text). The O(2) HF2 line is split by CDW order. }
\end{figure*}

$^{17}$O-enriched untwinned single crystals of YBa$_2$Cu$_3$O$_y$ were prepared with $y_{\rm nominal} = $~6.47 (ortho-II oxygen ordering, hereafter O-II), 6.56 (O-II), 6.68 (O-VIII) and 6.77 (O-III), following ref.~\cite{Liang12}. Their high quality is attested by the very sharp NMR lines, their oxygen order is attested by the observation of inequivalent chain or planar sites~\cite{Wu16} and their doping level is attested by the values of the superconducting, vortex melting and CDW transition fields/temperatures, which are all consistent with the literature~\cite{Wu13,Wu15}. More information about samples and experimental methods can be found in supplemental material~\cite{suppl} and in refs.~\cite{Wu13,Wu15,Wu16}.

Fig.~1a shows a typical NMR spectrum in the field-induced CDW state of YBa$_2$Cu$_3$O$_{6.56}$, for O(2) and O(3) sites, which are those sites in CuO$_2$ planes lying in bonds oriented along the crystalline $a$ and $b$ axes, respectively~\cite{suppl}. Each site comprises five lines corresponding to the different nuclear transitions of the nuclear spin $^{17}I=5/2$. Below $\sim$100~K, we observe that the separation between O(3E) lines (sites below empty chains) and O(3F) (below full chains) is much smaller than the linewidth and thus these sites are not distinguished. 

For both O(2) and O(3), the low-frequency satellites (LF1 and LF2) present a clear asymmetric profile with a long tail toward high frequencies. The same asymmetry is also clear on the first high-frequency (HF1) satellite of O(3) as well as on the central line. Other lines are either split by charge order (O(2) HF1 and HF2) or they experience strong quadrupole broadening (O(2) and O(3) LF2), which in both cases makes the asymmetry, if any, less visible. Detailed analysis of these spectra~\cite{suppl} indicates that the asymmetry is actually present on each individual line and that it does not result from an unresolved line splitting on some of the lines. The asymmetric line broadening actually adds to both the splitting produced by long-range CDW order~\cite{Wu11,Wu13} and the symmetric broadening due to short-range CDW order~\cite{Wu15}.

The main outcome of the analysis is that the asymmetry arises exclusively from a spatial distribution of local magnetic fields, not from a distribution of electric-field-gradients. The histogram of these local fields is directly given by the NMR lineshape, provided other CDW-induced effects are comparatively small (as is the case for LF1 lines, for example). Therefore, the triangle-shaped distribution points to an inhomogeneous state in which an overwhelming majority of sites experience small local fields whereas sites with large fields are relatively rare. This is the first central result of this work.

We quantify the asymmetry by fitting the lines with Gaussians having distinct right and left widths $w_R$ and $w_L$  ($w_R \geq w_L$). As shown in Fig.~2a,b, all of the asymmetry arises from the broadening of the right (high-frequency) part of the line, meaning that the distribution involves only enhanced values of the local field. Furthermore, both the field and the temperature dependence $w_R$ (or equivalently of the asymmetry $A=(w_R-w_L)/(w_R+w_L)$) closely follow the variation of the $^{17}$O line splitting that provided direct evidence of CDW order in high fields~\cite{Wu13}. Therefore, the spatial distribution of local fields arises only in conjunction with long-range CDW order. This is the second central result of this work.

The asymmetric boadening at $T=3$~K becomes field-independent above $\sim$25~T~(Fig.~2c,d). At these fields, the bigaussian with $A\simeq 0.35$ is actually very close to an extreme value distribution (EVD) function, $f(x)=b\exp(-\exp(-z)-z+1)$ with $z=(x-x_c)/w$, found in a variety of critical phenomena~\cite{EVD1,EVD2,EVD3,EVD4,EVD5,EVD6}. Quite remarkably, this asymmetry in high fields is found to be the same in four samples having not only different oxygen contents but mostly different levels of disorder, as indicated by the factor of 3 difference in their high temperature linewidth $w_{\rm HT}$~\cite{suppl}. The identical asymmetry for all samples simply follows from the fact that both $w_R$ and $w_L$ are proportional to $w_{\rm HT}$~\cite{suppl}. This strongly suggests that the asymmetric broadening is triggered by disorder. This is our third main observation. 

That the anomalous distribution is both tied to long-range CDW order disorder immediately suggests that it is unrelated to the vortex lattice or to magnetic order. This is confirmed by the following observations.
First, vortex broadening typically decreases with increasing field, contrary to what we observe here. Second, the distribution is observed up to at least 45~T while the vortex lattice melts near $B_{\rm melting}\simeq24$~T at $T=3$~K~\cite{Ramshaw12}. Therefore, in most of the field and temperature range investigated, vortices fluctuate much faster than our typical spectral resolution of $\sim$10~kHz and thus cannot broaden the NMR spectrum. For the same reason, any electronic pattern associated with the vortex cores (such as Andreev bound states) cannot explain the results. Last, if due to magnetic order, the broadening should be preceded by an enhancement of the relaxation rates, which is not seen in either $^{63}$Cu or $^{17}$O data. 

\begin{figure}[t!]
\centerline{\includegraphics[width=3.5in]{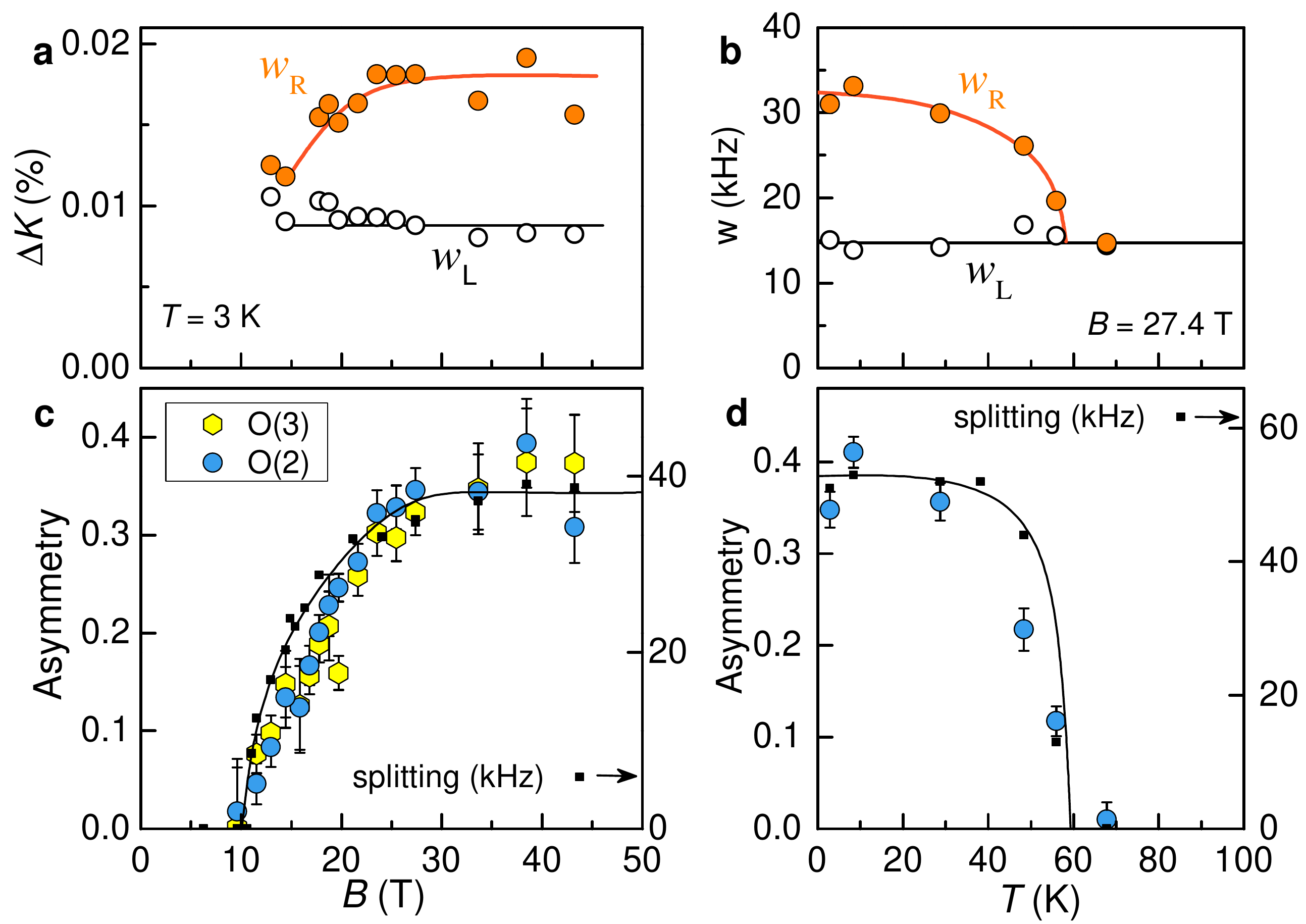}} 
 \caption{(Color online). (a,b) Field and temperature dependence of $w_{\rm R}$ and $w_{\rm L}$, the right (high-$f$) and left (low-$f$) width of the O(2) LF1 line in YBa$_2$Cu$_3$O$_{6.56}$. (c,d) Field and temperature dependence of the line asymmetry for O(2) and O(3) LF1 lines, compared to the quadrupole part of the O(2) HF2 line splitting $\Delta \nu_Q$ from ref.~\cite{Wu13}. The field was tilted off the $c$-axis toward $b$ in (a,c) and toward $a$ in (b,d). }
\end{figure}

Having excluded broadening by vortices and magnetic order, it is then appropriate to speak in terms of inhomogeneous Knight shift ($K$) of a paramagnetic metal. $K$ is defined as the shift $K=(f-f_0)/f_0$ of the  resonance frequency $f$ with respect to a reference  $f_0$, due to the local magnetic susceptibility. For $B\parallel \alpha$ where $\alpha=a,b,c$ represents the crystallographic axes of YBCO, the diagonal components of the Knight shift tensor $K$ are related to the static, uniform, spin susceptibility $\chi^{\rm spin} = \chi^{\rm spin}(q=0,\omega=0)$ through:
\begin{equation} 
K_{\alpha\alpha} (T) = K_{\alpha\alpha}^{\rm spin} (T)+ K_{\alpha\alpha}^{\rm orb} = \frac{A^{\rm hf}_{\alpha\alpha}}{g_{\alpha\alpha}\mu_B} \chi_{\alpha\alpha}^{\rm spin}+ K_{\alpha\alpha}^{\rm orb}
\end{equation}
where $A^{\rm hf}$ is the hyperfine tensor and $g$ the Land\'e factor. $K^{\rm orb}$ is mostly attributed to Van-Vleck paramagnetism. 

Although it must be related to CDW order in some way (Fig.~2), the shift distribution cannot be a direct imprint of the charge modulation. This is because the high-field CDW with period $\sim3b$~\cite{Gerber15,Chang16} has already been shown to produce a bimodal distribution of $K$, tied to the bimodal distribution of quadrupole frequencies $\nu_{\rm quad}$ that directly reflects the charge modulation. The total splitting is then contributed about equally by the $K$ and $\nu_{\rm quad}$ bimodal distributions~\cite{Wu11,Wu13}. In contrast, there is no $\nu_{\rm quad}$ distribution accompanying the skewed distribution of $K$ here~\cite{suppl}. Further, the continuous aspect and the skewness of the lineshape actually suggest that the local field is maximal at relatively few locations and that it decays over a typical distance much larger than a lattice step. There are not many examples of such Knight shift distributions. For instance, oxygen vacancies in the chains or impurities in the planes produce a staggered magnetization that leads to symmetric NMR broadening~\cite{Chen09,Julien00,Ouazi06}. 

\begin{figure}[b!]
\centerline{\includegraphics[width=3.3in]{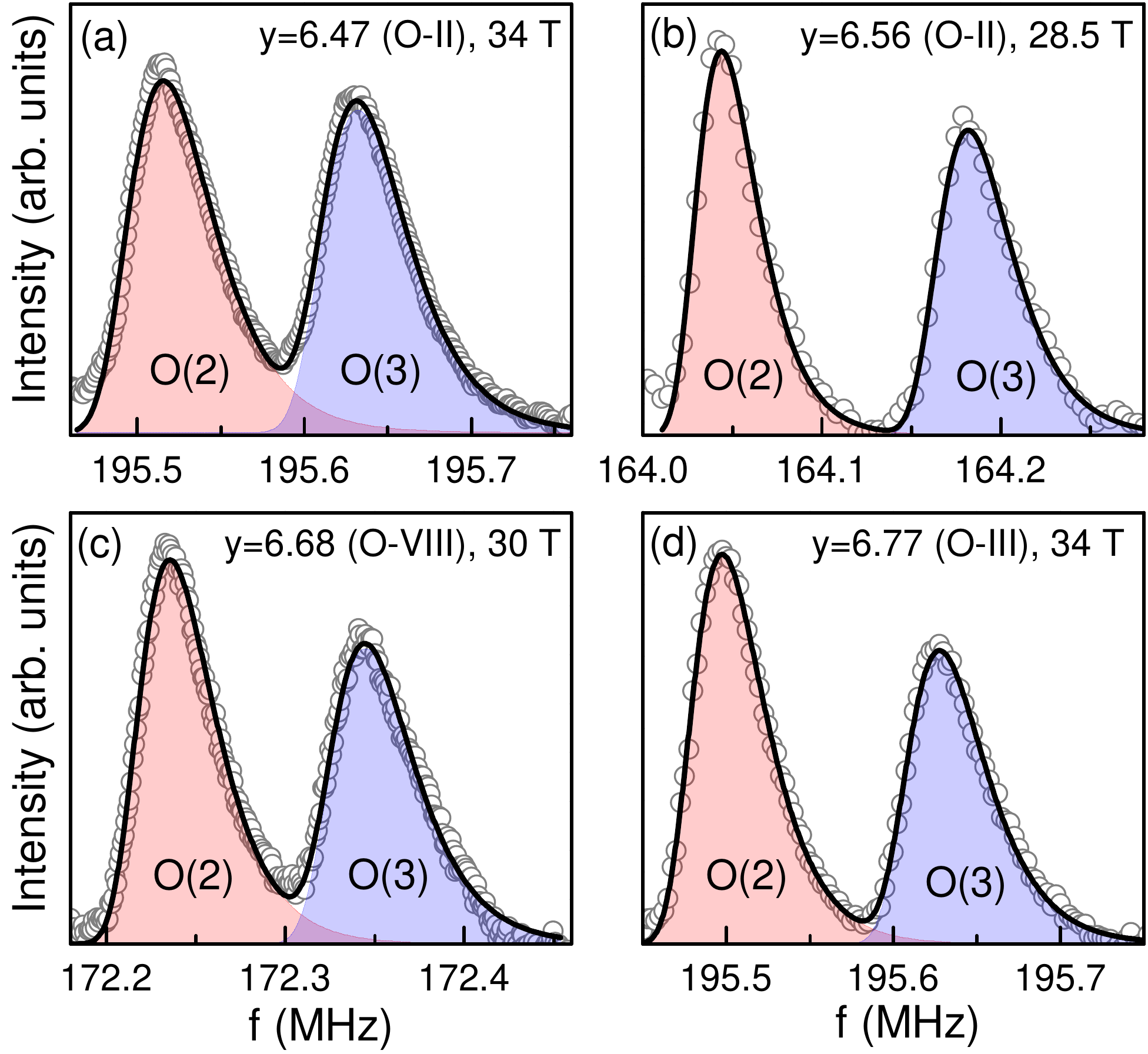}} 
 \caption{(Color online). O(2) and O(3) first low frequency quadrupole satellites showing similar asymmetric profile for four different samples (see methods). Continuous lines are fits of each peak to an extreme value distribution (see text).}
\end{figure}

\begin{figure}[t!]
\centerline{\includegraphics[width=3.5in]{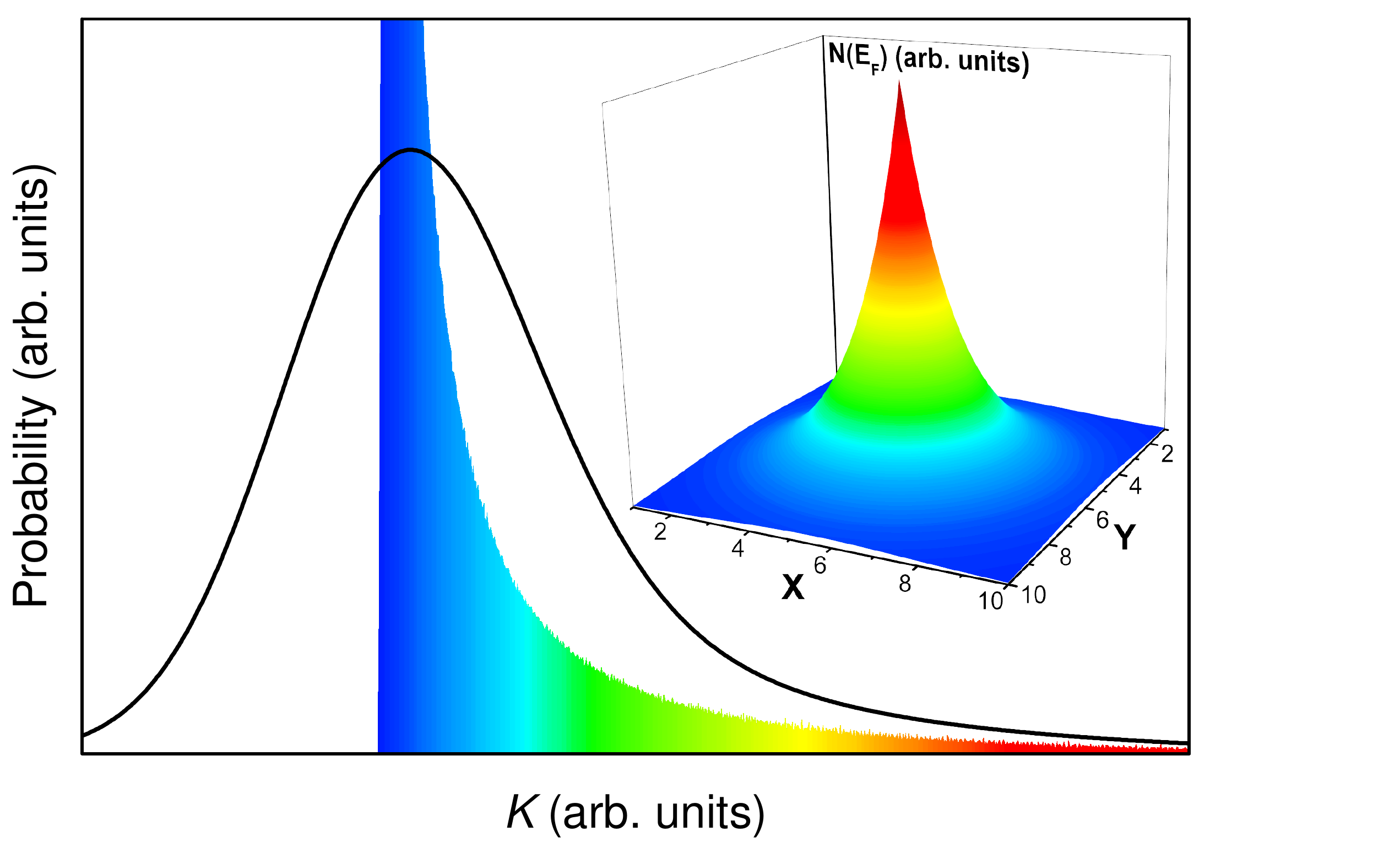}} 
 \caption{(Color online). Probability distribution of the Knight shift $K\propto N(E_F)$ for the real space pattern of $N(E_F)$ shown in inset (exponential decay around a given position in the $(x,y)$ plane). The black line is the histogram convoluted with a Gaussian distribution. }
\label{histogram}
\end{figure}

Actually, the observed skewed distribution ressembles the histogram of local density-of-states at the Fermi level, N$(E_F)$, in the presence of the so-called "impurity resonances" or "scattering resonances" observed by scanning tunneling microscopy (STM) around defects in the superconducting state of cuprates~\cite{Hudson99,Yazdani99,Pan00,Balatsky06,Alloul09}, or in their pseudogap state~\cite{Chatterjee08}. In Zn-doped YBa$_2$Cu$_3$O$_7$, these impurity states have been shown to produce the same asymmetric profile of $^{17}$O NMR lines as observed here~\cite{Ouazi06,Harter07}, with an asymmetry also independent of disorder (that is, of Zn content~\cite{suppl}). In a simple metal picture, $N(E_F)$ naturally enters into $\chi^{\rm spin}$, and thus into $K$ (Eq.~1). The broadening of the high-frequency side of the lines directly reflects the enhanced $N(E_F)$ and its decay away from the defects, as schematically shown in Fig.~\ref{histogram}. We are not aware of any other physical phenomenon giving rise to a similar distribution of $N(E_F)$, except for Anderson localization~\cite{Takagi81,Richardella10} that should be irrelevant here. We therefore conclude that the asymmetric broadening must be related to quasiparticle bound states around defects in the high-field CDW state. This is quite paradoxical: not only is YBCO a particularly clean cuprate but, instead of developing in the superconducting state, the effect grows here on quenching superconductivity with high fields. 

Impurity bound states have been argued to be a generic property of metals with Dirac-type electronic dispersion~\cite{Balatsky14}. In addition to the cuprates, they have been observed in iron-based~\cite{Yin15,Yang13} and heavy fermion~\cite{Zhou13} nodal superconductors as well as in graphene~\cite{graphene} and at the surface of topological insulators~\cite{TI}. With few exceptions (see below), they have not been observed in CDW materials in general, which makes our observation even more puzzling. 

Since neither the pseudogap (that indeed persists in the high field CDW~\cite{Wu13}) nor the substantial CDW modulations of the normal state (the correlation length $\xi_{ab}^{\rm CDW}$ reaches 20 lattice spacings) appear to be sufficient conditions for the formation of bound states here, it must be that one of these three characteristics of the field-induced CDW (or a combination thereof) is pivotal: the large values of $\xi_{ab}^{\rm CDW}$, the $c$-axis coherence ({\it i.e.} $\xi_{c}^{\rm CDW}/c> 1$) or the uniaxial nature of the field-induced CDW~\cite{Gerber15,Chang16}. Here, we find that $c$-axis coherence is unlikely to be a key player  since the asymmetric broadening (Fig.~2) starts at $H_{\rm charge}^{2D}\simeq 10$~T at which $\xi_{ab}^{\rm CDW}$ starts to grow significantly~\cite{Chang16} and not at $H_{\rm charge}^{3D}\simeq 17$~T that marks the onset of the growth of  $\xi_{c}^{\rm CDW}$~\cite{Chang16}.

That the apparition of bound states does not coincide with the onset of $c$-axis coherence also suggests that quasiparticles are not scattered directly off out-of-plane defects. However, chain-oxygen defects have been shown to provide the main source of electronic scattering~\cite{Bobowski10} and they are indeed ubiquitous in the chain layer of oxygen-ordered YBCO~\cite{Wu16}. Therefore, it is likely that bound states are still related to out-of-plane disorder rather than to in-plane impurities or vacancies. This paradox may be resolved if the chains are involved only indirectly in the scattering, that is, via electronic perturbations created in the planes by chain defects. Possible candidates are phase slips or amplitude defects of the CDW, Friedel oscillations~\cite{Demler15}, patches of uncondensed short-range CDW order~\cite{Wu15,Gerber15,Chang16} and patches of short-range spin-density-wave order (long-range CDW order triggers slow  fluctuations~\cite{Wu11} that could be pinned by disorder and create a Kondo resonance~\cite{Polkonikov01}). 

It should also be noted that impurity resonances have actually been observed in the chains of YBa$_2$Cu$_3$O$_{6.97}$~\cite{Derro02} and so one could speculate that we are seeing here the impact onto the planes of a similar effect. However, the apparition of the asymmetric broadening at $H_{\rm charge}^{2D}$ again makes this explanation implausible.

\begin{figure}[t!]
\centerline{\includegraphics[width=3.2in]{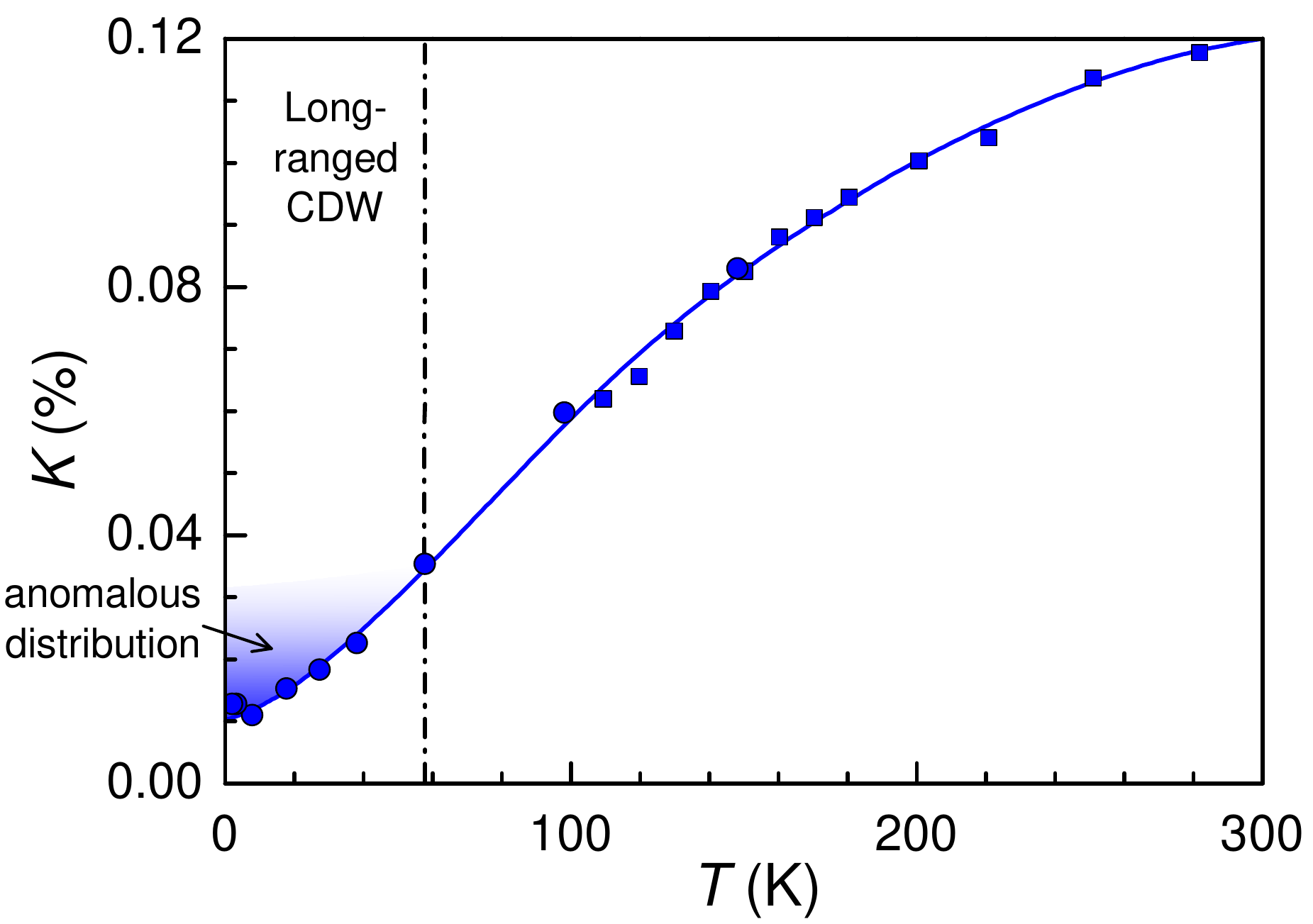}} 
 \caption{(Color online). Knight shift $K$ measured at 28.5~T (circles) and its asymmetric distribution in $B\gtrsim 20$~T, schematically represented by the colored area ($B \parallel c$, YBa$_2$Cu$_3$O$_{6.56}$). $K$ is defined as the most probable value of the distribution, {\it i.e.} at the position of maximum intensity on the line of O(2,3) sites. Squares are shift data measured at 12.0~T. Above 150 K, O(2) and O(3) central lines can be resolved so $K$ is defined as the average shift of the two lines. The line is a guide to the eye.}
 \label{shift}
\end{figure}

We see that, in addition to exposing a novel case of electronic bound states, our results raise a number of questions. In the following, we argue that these important questions provide an insightful window into charge ordering in the cuprates. 

First, identifying the nature of the defects responsible for scattering is an important task as all of the above-mentioned possibilities touch upon open unsolved aspects of the CDW in YBCO. The defects could be phase slips if the CDW incommensurability seen by X-ray scattering in high fields is due to the presence of discommensurations that separate domains of locally commensurate order~\cite{Mesaros16}. Whether the CDW is intrinsically incommensurate or not is a crucial question to unravel its microscopic origin. We note that in-gap states, attributed to solitonic defects, have been reported in one-dimensional CDW materials~\cite{Latyshev05,Kim12,Brun15}. Also important is the elucidation of the role of oxygen defects in the chain layer. This should shed light on a hypothetical role of the chains in the high-field CDW~\cite{Bozin16} as well as on the nature of pinning of the short-range CDW~\cite{Wu15,LeTacon14,Fukuyama78}. 

Clarifying which aspect of the CDW is crucial in the formation of bound states should be informative on its microscopic nature. A natural question is whether the effect is sensitive to the uniaxial nature of the modulation in high fields~\cite{Wu11,Chang16} or to a possible $d$-wave symmetry of the intra-unit-cell form factor~\cite{Fujita14,Comin15} in high fields. Theoretically, bound states have been found, under certain conditions, in models of one-dimensional CDW~\cite{Tutto85,Schuricht11} and in the $d$-density-wave state~\cite{Zhu01,Wang02,Morr02,Ghosal04,Nielsen12,Vanyolos07}, but they do not occur at $E_F$ (note that they only need to have a finite weight at $E_F$ to contribute to the Knight shift). Also, the possible presence of a Dirac cone in the band structure could provide clues on the reconstructed Fermi-surface in high fields.

Finally, it should be mentioned that recent proposal of a pair-density-wave (PDW) state in the cuprates~\cite{Agterberg15,Lee14,Fradkin15,Hamidian16} provides a tantalizing explanation of our data in terms of Andreev bound states at locations where the superconducting gap changes its sign. Zero-bias conductance peaks, a standard signature of bound states in STM, have been reported in La$_{1.88}$Sr$_{0.12}$CuO$_4$ and have actually been interpreted as evidence of PDW~\cite{Yuli10} (see also Refs.~\cite{Galvis13,Galvis14} for zero-bias conductance peaks in 2D chalcogenides). However, we observe bound states up to at least 45~T (Fig.~2), far above the upper critical field $H_{c2} \simeq 24 \pm$~2 T in YBa$_2$Cu$_3$O$_{6.56}$~\cite{Grissonnanche14,Grissonnanche16,Marcenat15}. In order to explain our results, Andreev bound states would then have to be pinned by disorder and to survive in the metallic but non-superconducting state above $H_{c2}$. While it has been envisaged that a phase-disordered~\cite{Agterberg08} or short-ranged fluctuating~\cite{Lee14} PDW state could survive above $H_{c2}$, none of these situations is likely to produce static bound states. Therefore, the PDW state does not readily explain our data.

Clearly, the results presented here call for theoretical investigations of the effects of disorder in two and three-dimensional CDW models for the cuprates. They should also stimulate further experimental work. In particular, a direct confirmation of the bound states, and possibly an identification of the defects, should come from STM experiments in high fields while some aspects of disorder pinning could be addressed by X-ray experiments~\cite{Ravy06}.

We thank W. Atkinson, D.~Agterbeg, A.~Balatsky, C.~Brun, J.C.~Davis, E.~Demler, P.~Hirschfeld, M. Hamidian, E.~Hudson, A.~Kampf, S.~Kivelson, P.~Lee, P.~Mallet, A. Mesaros, V.~Mitrovi\'c, J.-P.~Pouget, C.~Proust, C.~Renner, A.~Rosso, E.~da Silva Neto, E.~Dalla Torre, A.-M.~Tremblay, M.~Vojta, A.Yazdani for communications related to this work. Work in Grenoble was supported by the French Agence Nationale de la Recherche (ANR) under reference AF-12-BS04-0012-01 (Superfield), by the Laboratoire d’excellence LANEF in Grenoble (ANR-10-LABX-51-01) and by the Universit\'e J. Fourier - Grenoble (SMIng/AGIR). Part of this work was performed at the LNCMI, a member of the European Magnetic Field Laboratory (EMFL). A portion of this work was performed at the National High Magnetic Field Laboratory, which is supported by the National Science Foundation Cooperative Agreement No. DMR-1157490, the State of Florida, and the U.S. Department of Energy. Work in Vancouver was supported by the Canadian Institute for Advanced Research and the Natural Science and Engineering Research Council.

\clearpage
\newpage
\onecolumngrid

\section{SUPPLEMENTAL MATERIAL }

\begin{figure}[b!]
\centerline{\includegraphics[width=3in]{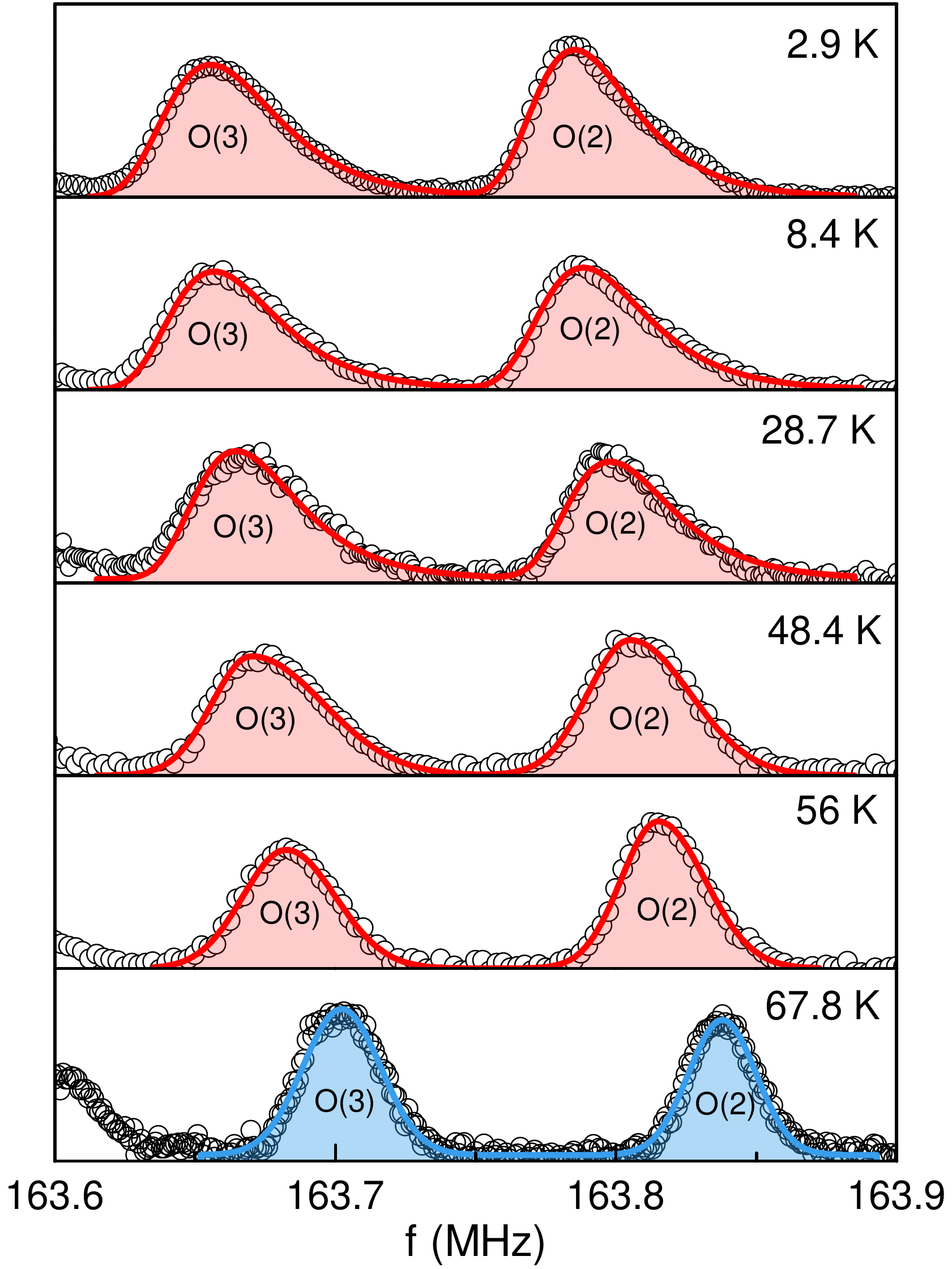}} 
 \caption{(Color online). Temperature dependence in $B=28.5$~T of the $^{17}$O(2) and $^{17}$O(3) first low frequency quadrupole satellites in YBa$_2$Cu$_3$O$_{6.56}$. Continuous lines are fits with an asymmetric Gaussian function ("bigaussian"). The field was tilted off the $c$-axis towards the $a$-axis.}
\end{figure}

\begin{figure*}[t!]
\vspace{-4cm}
\centerline{\includegraphics[width=8in]{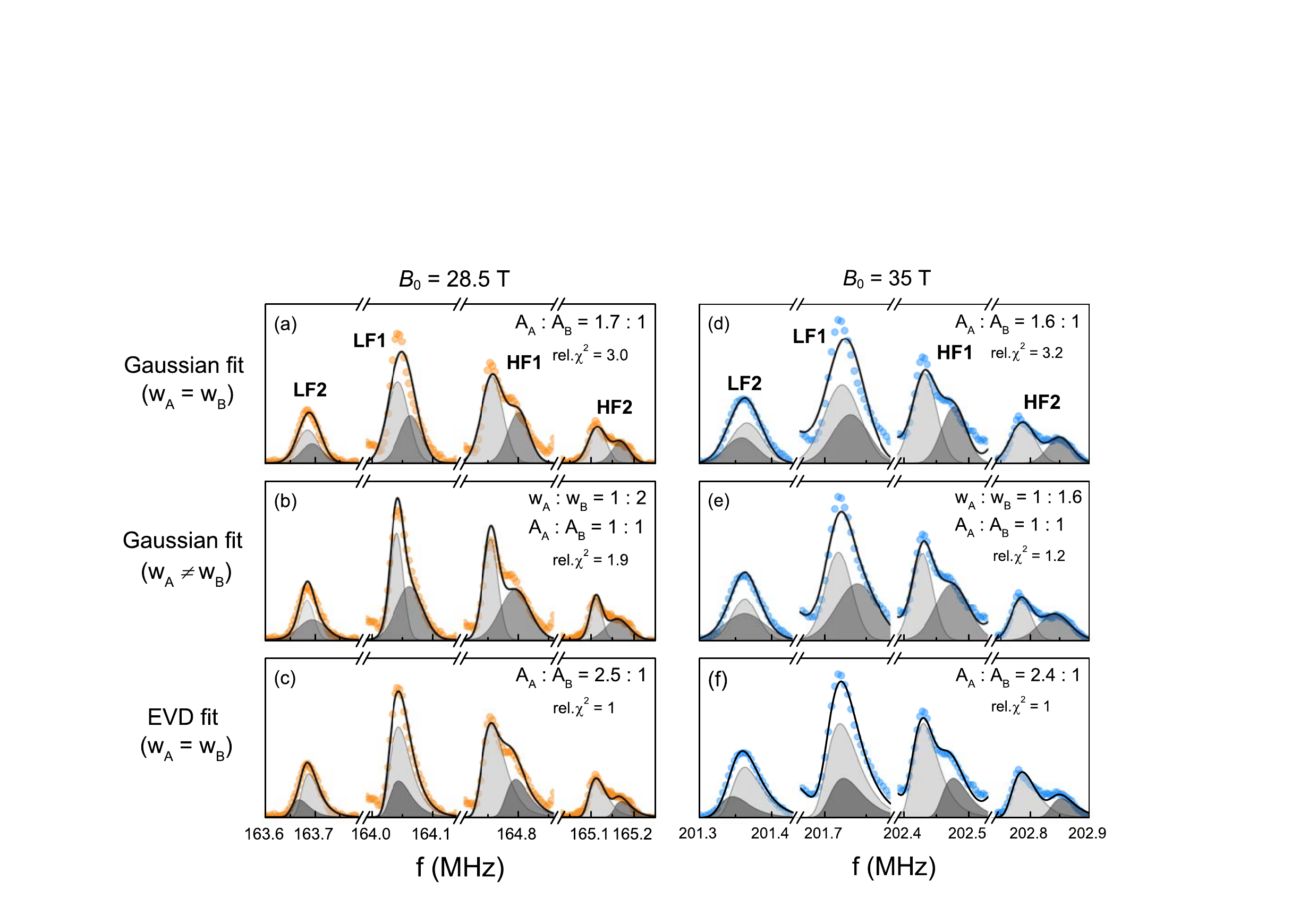}} 
 \caption{(Color online). $^{17}$O quadrupole satellites of O(2) in YBa$_2$Cu$_3$O$_{6.56}$ at 28.5 (left panels) and 35~T (right panels), together with fits within three different scenarios: (a,d) the line asymmetry is produced by an unresolved splitting and the CDW-induced splitting is described by two gaussians of equal width and unequal areas. (b,e) The line asymmetry is produced by an unresolved splitting and the CDW-induced splitting is described by two gaussians of unequal widths and equal area. (c,f) The line asymmetry is an intrinsic property of each individual line and the CDW-induced splitting is described by two asymmetric peaks having equal width and unequal areas.}
  \label{satellitefits}
\end{figure*}

\subsection{Experimental methods} 

Fields $B$ above 20~T and 35~T were provided by resistive magnets in Grenoble and by the hybrid magnet in Tallahassee, respectively. Samples were mounted in a goniometer with their crystalline $c$-axis tilted off the $B$ direction (toward the $b$-axis, unless otherwise stated) by an angle $\theta\simeq$16-18$^{\circ}$ so as to separate O(2) and O(3) lines while keeping a large component of $B$ along $c$~\cite{Wu13}. In general, we quote values of the component of $B$ along $c$, except for plots of the NMR spectra for which  total values of $B$ are quoted. 

\begin{figure*}[t!]
\centerline{\includegraphics[width=3.7in]{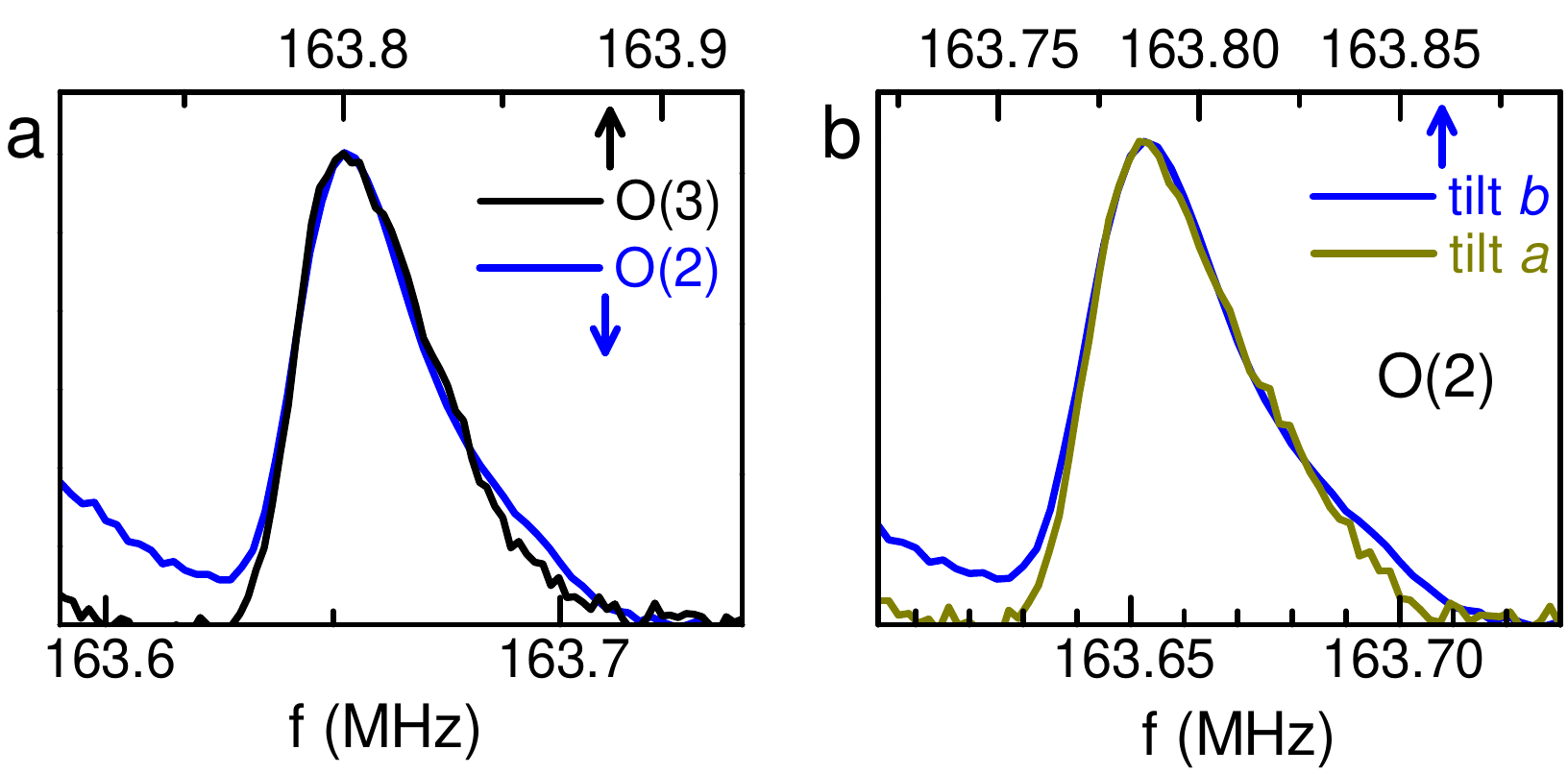}} 
 \caption{(Color online). (a) LF1 lines for $^{17}$O(2) and $^{17}$O(3) sites. (b) LF1 lines for $^{17}$O(2) with the field tilted by 18$^\circ$ away from the $c$-axis in the $a$ and in the $b$ direction. All data correspond to $B=$~28.5~T and $T=$~3~K in YBa$_2$Cu$_3$O$_{6.56}$.}
 \label{robust}
\vspace{1cm}
\centerline{\includegraphics[width=6in]{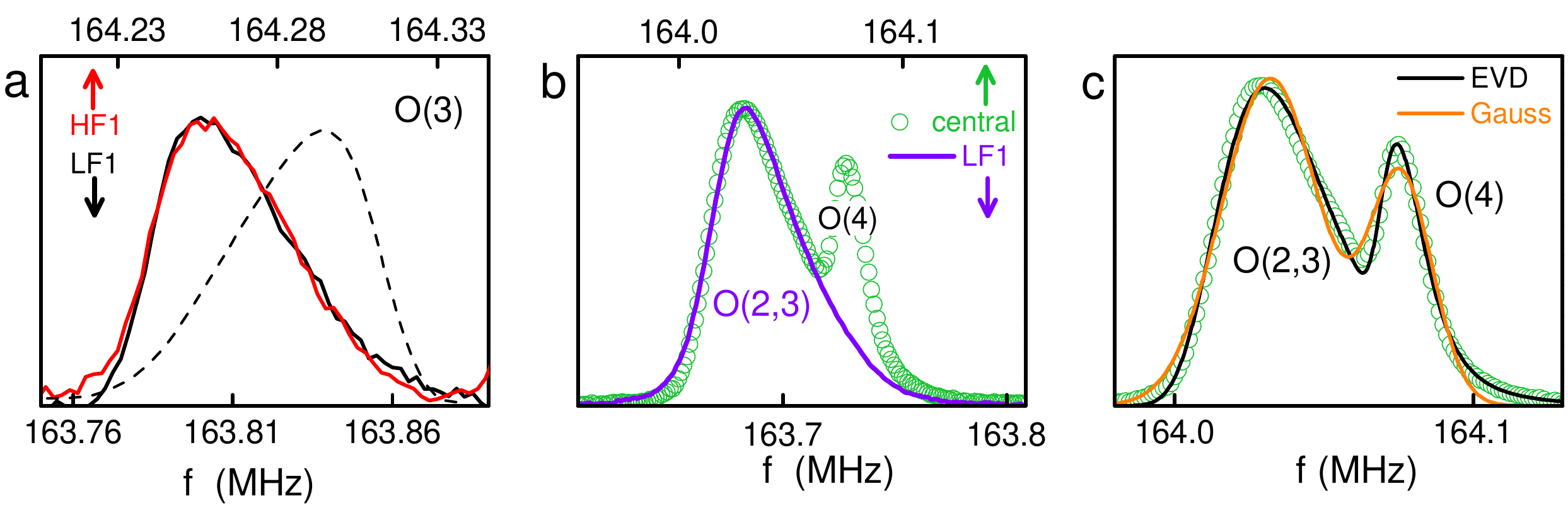}} 
 \caption{(Color online). (a) Exact scaling of LF1 and HF1 satellites of O(3). The dashes show the opposite asymmetry that one of the two lines would present if the skewed distribution was of quadrupolar origin. (b) The O(2,3) LF1 line showing the same asymmetric profile as the O(2,3) central line (O(2) and O(3) overlap because the field was exactly aligned with the $c$-axis in this particular set of data). (c) Fits of O(2,3) and apical O(4) central lines with Gaussians and EVD functions, the second being considerably better. All data correspond to $B=$~28.5~T and $T=$~3~K in YBa$_2$Cu$_3$O$_{6.56}$.}
 \label{o3}
\end{figure*}
\subsection{Analysis of $^{17}$O(2) spectra} 

The main question that arises in interpreting NMR spectra of the O(2) site is whether the asymmetric broadening of the lines is an intrinsic feature of all individual lines or whether it corresponds to a CDW-induced splitting that is unresolved for some of the lines. 

Indeed, CDW-induced splittings are clearly resolved on the high-frequency satellites HF1 and HF2 (for which modifications in the hyperfine magnetic shift and in the quadrupole frequency are additive) but not on the low-frequency satellites LF1 and LF2 (for which magnetic and quadrupole effects are subtractive and therefore approximately compensate each other)~\cite{Wu13}. See Fig.~\ref{satellitefits}. It then appears legitimate to suspect that magnetic and quadrupole splittings do not exactly compensate each other on LF1 and LF2 (especially as they cannot compensate exactly on LF1 and LF2 simultaneously), so that this residual splitting appears as an asymmetric profile. In order to rule out this scenario, it is desirable to show that the asymmetry is also present on the CDW-split satellites HF1 and HF2 lines. However, the strong overlap of the CDW-split lines masks the asymmetry.

These observations call for a detailed fitting of the lineshapes across the whole spectrum. The main difficulty, however, is that the adequate model with which to fit the $^{17}$O NMR lineshapes is not known. 

According to high-field X-ray data, the CDW pattern is unidirectional (propagating along the $b$ axis), with an incommensurate period $\lambda\simeq(3.2\pm0.1)b$~\cite{Chang16}. In principle, a unidirectional (1$q$) incommensurate sinusoidal wave produces a two-horn NMR lineshape that, after convolution by gaussian broadening, looks like two overlapping peaks of equal intensities. In contrast, our $^{17}$O HF1 and HF2 lines clearly show two peaks of unequal intensity~\cite{Wu13}. Explaining this intensity difference is beyond the scope of this paper, so we performed phenomenological fits using two inequivalent sites, that is, two sets of satellite peaks. Attempts to fit with three inequivalent sites proved unsuccessful. 

The fits were performed simultaneously on the four quadrupole satellites of $^{17}$O by imposing that each set of peaks has resonance frequencies $f$ such that  $\left( {{f}_{\text{HF2}}}-{{f}_{\text{HF1}}} \right):\left( {{f}_{\text{HF1}}}-{{f}_{\text{LF1}}} \right):\left( {{f}_{\text{LF1}}}-{{f}_{\text{LF2}}} \right)=1:2:1$, as dictated by the $I=5/2$ quadrupole spectrum. The width ratio of the two sets of peaks was either fixed to 1 (Fig.~\ref{satellitefits}a,d) or to the value given by fitting HF1 and HF2 satellites only [Fig.~\ref{satellitefits}(b,e)]. In all cases, the area ratio was fixed to the value obtained from a fit to HF1 and HF2 only. 

As Fig.~\ref{satellitefits} shows, fits using two sets of asymmetric peaks (each described by an extreme-value-distribution function) are globally better than those using Gaussian distributions (relative $\chi^2$ values are shown in the figure) and the obtained areas are in a ratio close to 2:1. The best Gaussian fits are obtained with two peaks of different width (typically 1:2 ratio) and 1:1 intensities. The quality of the fits differs mostly on LF1 satellites. For HF1 and HF2 transitions, on the other hand, fits are about equally good with and without asymmetric profiles (Fig.~\ref{satellitefits}). These fits are quite suggestive of an intrinsic asymmetry of each individual O(2) line but they cannot be considered as a conclusive proof of it. 

\subsection{Analysis of $^{17}$O(3) spectra} 

O(3) lineshapes provide a more compelling evidence that the asymmetric line profile does not result from a residual splitting. There are two main arguments that do not rely on fitting:

(1) The asymmetry turns out to be extremely robust: it is identical for O(2) and O(3) sites (Fig.~\ref{robust}a), identical for $H$ tilted either towards $a$ or $b$ axes (Fig.~\ref{robust}b) and identical in different samples having quite different line widths (Fig.~3 of main paper). Therefore, although a variety of experimental situations has been explored, no shoulder or bump suggestive of an underlying splitting has ever been detected on any of the low frequency satellites; instead a smooth profile and the same asymmetry value in high fields have been observed in all cases. 

(2) HF1 and LF1 satellites of the O(3) site present exactly the same asymmetric profile with a long tail towards high frequencies [Fig.~\ref{o3}a]. This observation implies that the asymmetry is purely a Knight shift distribution, namely it cannot involve any quadrupole effect (that appears to be vanishingly small on this line when the field is tilted off the $c$-axis towards the $b$-axis). One could still argue that the asymmetry is produced by a residual magnetic splitting ({\it i.e.} a small Knight shift difference between two unresolved sites) but this is extremely unlikely because the Knight shift of O(3) is  insensitive to the CDW. This is true for the short-range CDW order in the normal state~\cite{Wu15}, and this is also true here in high fields where quadrupole modifications are seen on both the LF2 and HF2 satellites of O(3), with relatively small difference between them, in contrast with the O(2) case where LF2 and HF2 are very different due to combination of quadrupole and magnetic splittings~\cite{Wu13}. Furthermore, it is quite improbable that a magnetic splitting produces exactly the same asymmetric broadening on O(3) as a magnetic and quadrupole splitting on O(2). 

\begin{figure*}
\vspace{1cm}
\centerline{\includegraphics[width=6in]{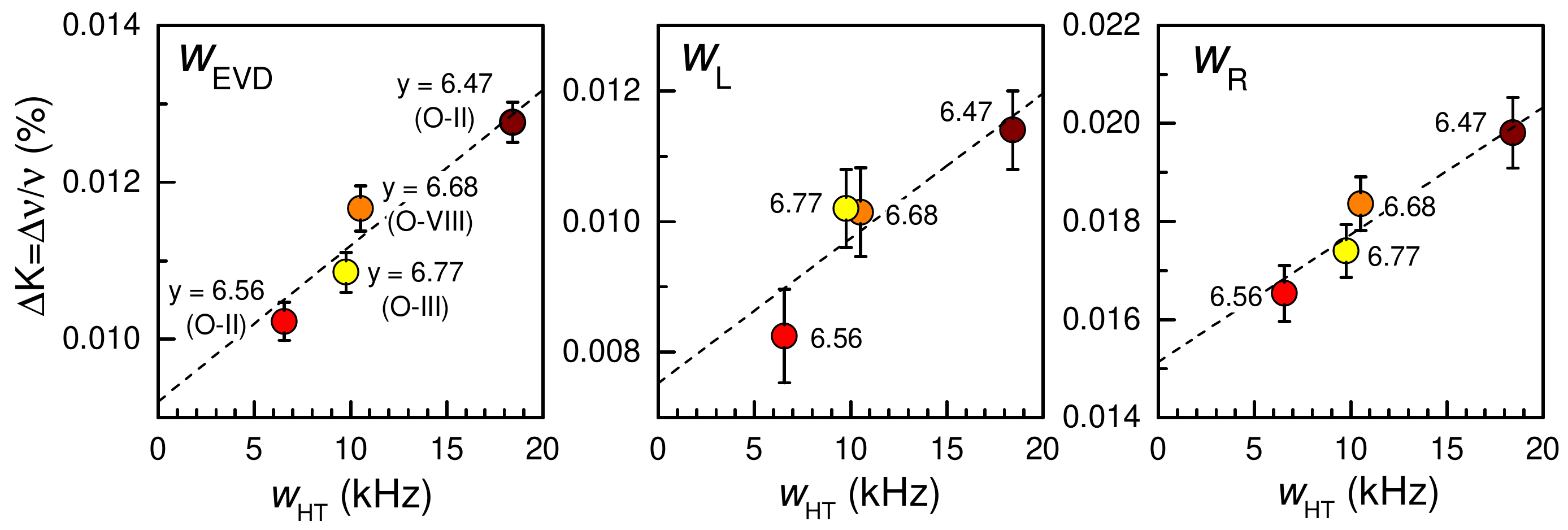}} 
 \caption{(Color online). Proportionality between the $^{17}$O high field/low $T$ broadening and the amount of disorder. The former quantity is represented by the width parameter of the EVD function ($w$ in the text; left panel), the left width $w_L$ contributed by the sample inhomogeneity and the CDW (middle panel) and the anomalous right width $w_R$ contributed by the bound states in addition to sample inhomogeneity and CDW (right panel). The data are taken at $T=3$~K and are expressed in terms of Knight shift distribution because they are taken at different fields ($B=28-34$~T). The latter quantity (disorder or sample inhomogeneity) is represented by the high temperature full width $w_{\rm HT}$ at $T=230-250$~K and $B=15$~T. All data are for the O(2EF), {\it i.e.} O(2) sites bridging Cu(2E) and Cu(2F) sites. Because both $w_L$ and $w_R$ are proportional to $w_{\rm HT}$, the line asymmetry at high fields ({\it i.e.} far above the CDW onset field) is identical in the four samples.}
\end{figure*}

\begin{figure*}
\vspace{1cm}
\centerline{\includegraphics[width=6in]{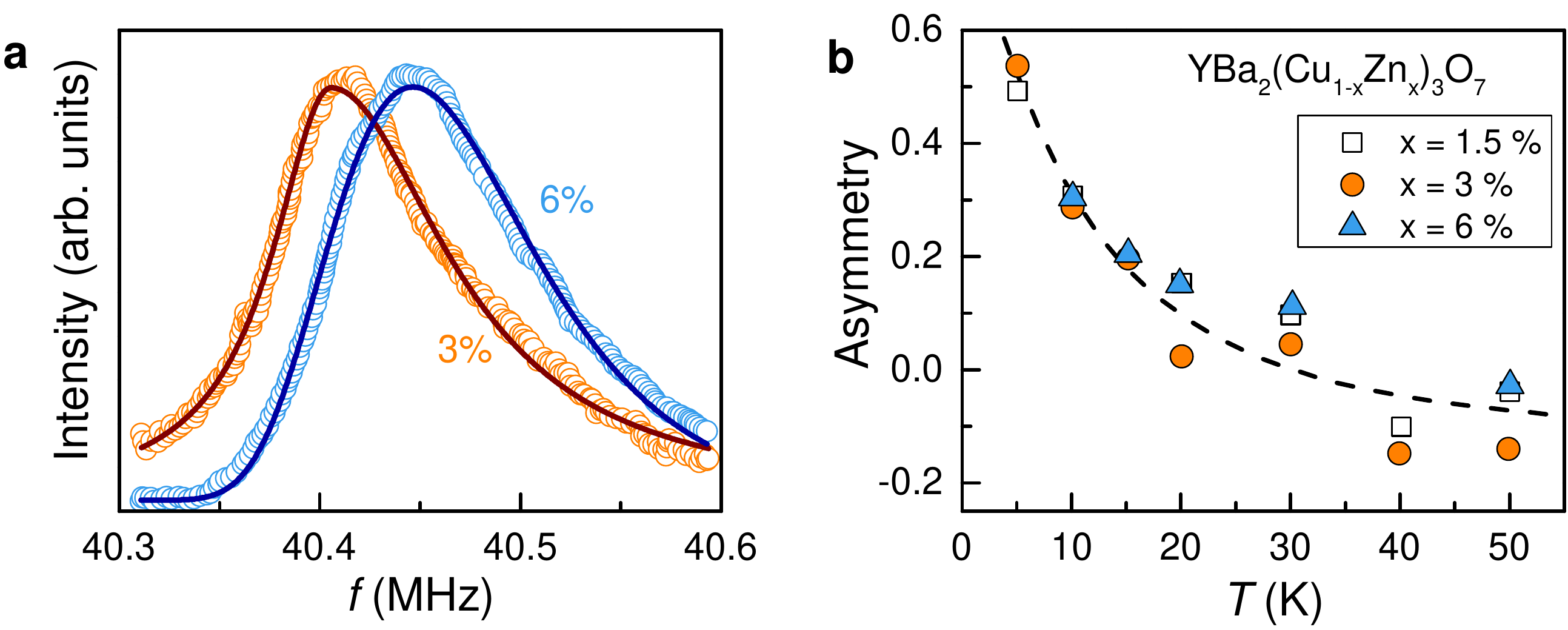}} 
 \caption{(Color online). (a) $^{17}$O(2,3) central lines in YBa$_2$Cu$_3$O$_{7}$ doped with $x$\% of Zn impurities from ref.~\cite{Ouazi06} at $B=$~7~T and $T=15$~K. Lines correspond to fit to an extreme value distribution of width $w\simeq50$~kHz ($x=6$\%) and to an asymmetric Lorentzian with $w_R/w_L=2$ ($x=3$\%). (b) Asymmetry (as defined in the text) of the $^{17}$O(2,3) central lines in YBa$_2$Cu$_3$O$_{7}$ doped with $x$\% of Zn impurities, calculated from data in ref.~\cite{Ouazi06}. While the total line width depends on the impurity concentration, the asymmetry does not. }
\end{figure*}

\subsection{Remark on hypothetical magnetic order}

Even though we have argued that static fields produced by frozen moments or by orbital currents are unlikely to explain our results, we discuss, for the sake of completeness, the requirements that putative ordered moments would have to fulfill in order to produce such an asymmetric broadening: i) their magnitude should be tied to the CDW order parameter (otherwise one could not explain the correlation with the line splitting in Fig.~2 of main paper). ii) There should be a ferromagnetic-like component pointing in the same direction for all of the oxygen sites in the two planes of each CuO$_2$ bilayer (otherwise, the broadening would not be seen only on the high-frequency side of the lines). iii) In order to produce a skewed field distribution, the moment magnitude should be spatially inhomogeneous with a strongly non-sinusoidal profile. Also, the absence of discernible peaks suggests a continuous distribution produced by a large number of inequivalent sites, that is, a long or incommensurate period. i) is theoretically conceivable (see ref.~\cite{Agterberg15}), while ii) and iii) could be realized, for instance, by canted moments of an in-plane spin-density wave (SDW).

\end{document}